\documentclass[]{aa}
\usepackage{psfig}
\def\ueber#1#2{{\setbox0=\hbox{$#1$}%
  \setbox1=\hbox to\wd0{\hss$ #2$\hss}%
  \offinterlineskip
  \vbox{\box1\box0}}{}}

\def\lesssim{\,\lower 1mm \hbox{\ueber{\sim}{<}}\,}
\def\grsim{\,\lower 1mm \hbox{\ueber{\sim}{>}}\,}

\makeatletter 
\let\@internalcite\cite
\def\cite{\@ifstar{\citeyear}{\citefull}}
\def\citefull{\def\astroncite##1##2{##1 ##2}\@internalcite}
\def\citeyear{\def\astroncite##1##2{##2}\@internalcite}
\def\citeau{\def\astroncite##1##2{##1}\@internalcite}
\def\citen{\def\astroncite##1##2{##1 (##2)}\@internalcite}
\def\possesivcite{\def\astroncite##1##2{##1's (##2)}\@internalcite}
\def\@citex[#1]#2{\if@filesw\immediate\write\@auxout{\string\citation{#2}}\fi
  \def\@citea{}\@cite{\@for\@citeb:=#2\do
    {\@citea\def\@citea{; }\@ifundefined
       {b@\@citeb}{{\bf ?}\@warning
       {Citation `\@citeb' on page \thepage \space undefined}}%
{\csname b@\@citeb\endcsname}}}{#1}}
\def\@cite#1#2{#1\if@tempswa , #2\fi}
\def\@biblabel#1{}
\makeatother

\begin{document}


\title{On the White Dwarf distances to Galactic Globular Clusters}

\author{M. Salaris\inst{1,2},
        S. Cassisi\inst{3,2},    
        E. Garc\'\i a--Berro\inst{4,6}, 
        J. Isern\inst{5,6}, 
        S. Torres\inst{4}}

\institute{Astrophysics   Research  Institute,   Liverpool  John  Moores
	   University,  Twelve Quays House,  Egerton  Wharf,  Birkenhead
	   CH41 1LD, UK 
	   \and
	   Max-Planck-Institut             f\"ur            Astrophysik,
	   Karl-Schwarzschild-Stra{\ss}e 1, D-85741 Garching, Germany
	   \and
	   Osservatorio  Astronomico  di  Collurania,  via  M.  Maggini,
	   I-64100 Teramo, Italy
	   \and 
	   Departament de F\'\i sica Aplicada, Universitat Polit\`ecnica
	   de  Catalunya,  c/Jordi  Girona   Salgado s/n,  M\'odul  B-4,
	   Campus Nord, 08034 Barcelona, Spain
	   \and 
	   Institut de Ci\`encies de l'Espai (CSIC), Edifici Nexus, Gran
	   Capit\'a 2--4, 08034 Barcelona, Spain
	   \and
	   Institut d'Estudis Espacials de Catalunya}

\offprints{M.~Salaris; (e-mail: ms@astro.livjm.ac.uk)}
\mail{M.~Salaris}

\date{Received; accepted}

\authorrunning{M. Salaris et al.}
\titlerunning{White dwarf distances to globular clusters}

\abstract{
We analyze in detail various  possible  sources of systematic  errors on
the distances of globular  clusters  derived by fitting a local template
DA white dwarf  sequence  to the  cluster  counterpart  (the  so--called
WD--fitting  technique).  We find  that  the  unknown  thickness  of the
hydrogen layer of white dwarfs in clusters plays a non negligible  role.
For  reasonable  assumptions  --- supported by the few sparse  available
observational  constraints  --- about the unknown mass and  thickness of
the hydrogen layer for the cluster white dwarfs, a realistic estimate of
the systematic error on the distance is within  $\pm$0.10 mag.  However,
particular  combinations of white dwarf masses and envelope  thicknesses
--- which at  present  cannot be  excluded  a priori  --- could  produce
larger  errors.  Contamination  of the  cluster DA  sequence  by non--DA
white dwarfs  introduces a very small systematic  error of about $-0.03$
mag in the $M_{\rm  V}/(V-I)$ plane, but in the $M_{\rm  V}/(B-V)$ plane
the systematic  error amounts to  $\sim\,+0.20$  mag.  Contamination  by
white dwarfs with helium  cores  should not  influence  appreciably  the
WD--fitting   distances.  Finally,  we  obtain  a   derivative   $\Delta
(m-M)_{\rm V}/\Delta  E(B-V)\sim\,-5.5$  for the WD--fitting  distances,
which is very  similar  to the  dependence  found  when  using  the Main
Sequence fitting technique.
 }

\maketitle

\section{Introduction}

Globular cluster stars are possibly the most suitable objects to be used
in order to constrain  the age of the  universe.  Since stars in a given
globular  cluster  are  coeval  and  share  the  same  initial  chemical
composition, the turn--off  brightness derived from the globular cluster
colour-magnitude  diagram (CMD) provides  straightforwardly  the cluster
age: if the  cluster  distance  is known,  one only has to  compare  the
observed cluster turn--off  brightness with the  correspondent  quantity
predicted  by  theoretical  isochrones  with  the  appropriate  chemical
composition in order to derive a firm estimate of its age.

The globular  cluster  distance scale is however still the subject of an
intense debate.  Main Sequence (MS) fitting  distances  using metal poor
Hipparcos  subdwarfs  with  well  determined   parallaxes  provide  long
distances  implying  globular  cluster ages of the order of 12 Gyr (see,
e.g.,  Gratton et  al.~1997).  The same long  distances  are  derived by
Salaris \& Weiss (1997,  1998),  Mazzitelli,  D'Antona \& Caloi  (1995),
Cassisi et al.~(1999) when using as standard  candles  theoretical  Zero
Age  Horizontal  Branch  (ZAHB)  models,  while  shorter  distances  are
obtained from the ZAHB models of Vandenberg et al.~(2000).  On the other
hand, the  calibration of the RR~Lyrae stars absolute  brightness  using
the statistical parallax methods (see, for instance, Fernley et al.~1998
and Luri et al.~1998)  provides  consistently much shorter distances and
therefore  higher  globular  cluster ages.  Taking these results at face
value, the uncertainty of the globular cluster distance scale appears to
be still of the  order  of  0.25--0.30  mag,  which  translates  into an
indetermination  on the  globular  cluster age of about 20\% --- see the
discussion in Renzini et al.  (1996).

An alternative empirical method to derive globular cluster distances has
been recently applied to the galactic globular cluster NGC~6752 (Renzini
et al.~1996).  The distance  indicator is in this case a template  local
sequence of white dwarf stars with effective temperature ($T_{\rm eff}$)
ranging   between   10\,000  and   20\,000  K,  and   precise   parallax
measurements,  which is fitted to the  dereddened  cluster  white  dwarf
sequence.  The vertical shift applied to the local template  sequence in
order to fit the sequence of the cluster provides its distance  modulus.
We shall refer to this technique as the WD--fitting  technique,  because
of its close analogy with the MS--fitting technique.

A key  assumption  of this method is that the white  dwarfs of the local
template  sequence --- which, in the case of the paper by Renzini et al.
(1996), have an estimated  average mass $M=0.515 \,  M_{\odot}$  --- are
totally  equivalent  to the white dwarfs of the cluster.  If this is the
case,  the  main  advantage  of  the  WD--fitting   technique  over  the
MS--fitting one is that it is in principle  independent of the knowledge
of the initial chemical  composition of the globular  cluster, since all
white dwarfs have  virtually  metal free  atmospheres,  and therefore it
avoids the  uncertainties  introduced in the  MS--fitting  method by the
colour  corrections  that have to be  necessarily  applied  to the local
subdwarfs in order to precisely  match the  metallicity  of the globular
cluster under scrutiny.

Another  potential  advantage of the WD--fitting  method with respect to
the MS--fitting  technique is, according to Renzini et al.  (1996), that
local  white dwarfs are more  abundant  than metal poor  subdwarfs,  and
therefore it is possible in principle  to have a larger  sample of local
calibrators.  However,  since  white  dwarfs  are much  dimmer  than the
subdwarfs used in the MS--fitting technique, up to now it is possible to
apply  the  WD--fitting  technique  only  through  HST  observations  of
relatively close globular clusters.

Finally it should be mentioned  as well that in the $T_{\rm  eff}$ range
considered by Renzini et al.  (1996),  local white dwarfs  appear in two
types,   either  as  the   so--called   DA  spectral   type,  which  are
characterized  by an envelope  made up of pure H (on top of a He layer),
or as the  non--DA   type, which is  characterized  by an almost pure He
envelope  and  possibly,  at least  for 20\% of them,  traces  of H with
abundances of the order of $10^{-4}$ by number, or even less --- see the
review by Koester \&  Chanmugam  (1990),  and  references  therein.  For
local white dwarfs the number ratio of DA versus non--DA is of the order
of 4:1 in this temperature  range.  Thus, the heterogeneity of the local
sample of white dwarfs could  potentially have undesired  effects on the
determinations of the ages and distances of globular  clusters using the
WD--fitting method.

Because of the potential  advantages offered by the WD--fitting  method,
we analyze in detail the possible systematic  uncertainties  involved in
this technique due to the poor knowledge of parameters affecting the CMD
location of the template and globular  cluster white  dwarfs,  expanding
upon  the  discussion  by  Renzini  et al.  (1996).  We  pay  particular
attention  to the role played by  differences  in the  thickness  of the
surface H and He layers on the final  distance  determination  --- which
has not yet been explored ---, to the possible differences between white
dwarfs in clusters with  predominantly  blue or red Horizontal  Branches
(HB), and also to the  differences  obtained  when  using DA or  non--DA
model atmospheres, or evolutionary models of white dwarfs with He cores.
All along our  analysis  we use  evolutionary  white  dwarf  models in a
purely differential way.

The paper is organized as follows.  In Sect. 2 we briefly describe the
theoretical  models,  while in Sect. 3 the  influence  of the  various
parameters affecting the CMD of white dwarfs is assessed.  An exhaustive
estimate of the systematic errors involved in the WD--fitting  technique
is derived in Sect. 4, which is followed by Sect. 5 were we draw our
conclusions.

\section[]{The models}

The stellar  evolution  computations  have been performed using the same
code and the same input physics  described in Salaris et al.~(2000).  We
just  recall  here that in the  temperature  range we are  dealing  with
($T_{\rm  eff}$  well  above  6\,000  K) the  OPAL  radiative  opacities
(Iglesias  \& Rogers  1993)  with  $Z=0$  are  used,  together  with the
conductive  opacities by Itoh et al.~(1983)  supplemented by the Hubbard
\& Lampe (1969) ones; the boundary conditions for the integration of the
stellar  structure  have been derived in this case from a grey $T(\tau)$
relationship,  which is completely  adequate at these  temperatures  ---
see,  for  instance,  Hansen  (1999)  and  Salaris  et  al.~(2000).  The
equation of state for the H and He envelopes is that of Saumon, Chabrier
\& Van Horn (1995),  while the  equation of state for the  carbon-oxygen
(CO) core is from Segretain et al.~(1994);  neutrino  energy losses have
been  taken  from  Itoh  et  al.~(1996).  Bolometric   luminosities  and
effective  temperatures  have been  transformed  into $V$ magnitudes and
colours  using the  relationships  by  Bergeron,  Wesemael \&  Beauchamp
(1995).

We have  computed  white dwarf model  sequences  for several  masses, CO
stratifications   ---  unless   stated   otherwise   our   reference  CO
stratification  is that from  Salaris et  al.~(1997)  --- and  various H
and/or He envelope thicknesses, which will be described in the following
section.  For all the computations an initial model with the selected CO
profile and envelope structure was converged at $\log(L/L_{\odot})  \sim
2.0$ and evolved down to sufficiently low temperatures.

\section{The white dwarf sequence location in the CMD}

As  already  mentioned  in \S1, the  WD--fitting  method is based on the
fitting of a template white dwarf sequence to the correspondent globular
cluster one.  The key ingredient to derive reliable distances (and hence
ages) is to ensure that the local white dwarfs  included in the template
sequence are homogeneous with the globular cluster ones.

\begin{figure}
\psfig{figure=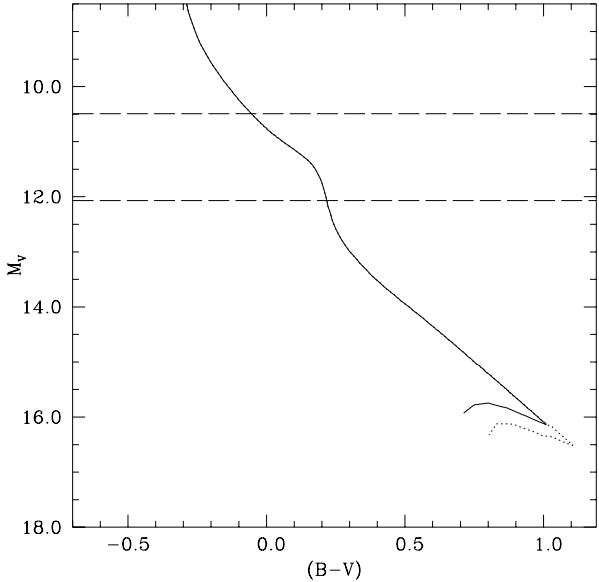,width=8.3cm,clip=}
\caption[]{Colour  magnitude  diagram of two cluster white dwarf cooling
sequences  with ages $t=10$  (solid line) and 12 Gyr (dotted  line), and
solar  metallicity, for the white  dwarf  progenitors  from  Salaris  et
al.~(2000).  The  horizontal  dashed  lines mark the region with $T_{\rm
eff}$ between 20\,000 and 10\,000 K.}
\end{figure}

\begin{figure}
\psfig{figure=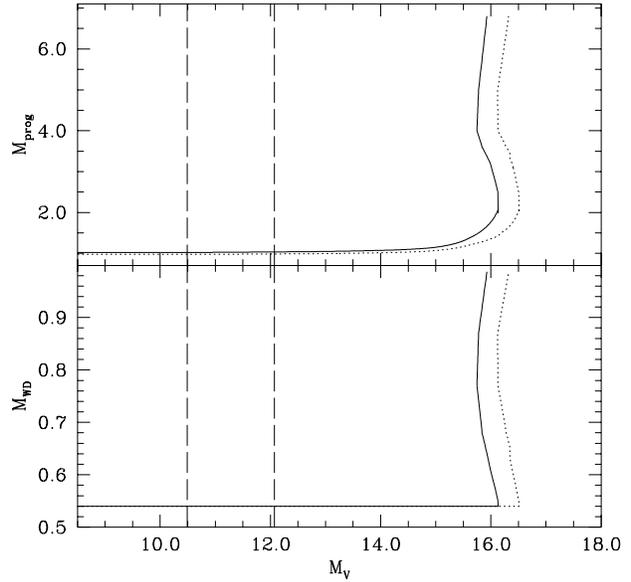,width=8.3cm,clip=}
\caption[]{Progenitor  mass  (upper  panel) and white  dwarf mass (lower
panel) as a function of $M_{\rm V}$ for the  cluster  cooling  sequences
displayed in Figure 1.  As in the previous  figure the  vertical  dashed
lines mark the region with $T_{\rm eff}$ between 20\,000 and 10\,000 K.}
\end{figure}

All stars in a globular  cluster are coeval;  therefore,  the luminosity
(and,  thus,  $T_{\rm  eff}$)  of a  white  dwarf  of a  given  mass  is
constrained by the fact that the sum of its cooling age ($t_{\rm cool}$)
plus the evolutionary time of its progenitor ($t_{\rm prog}$) --- which,
to a good  approximation, is equal to its main sequence  lifetime, since
subsequent evolutionary phases are much shorter --- must be equal to the
globular  cluster age ($t_{\rm GC}$).  It is immediately  clear that one
also has to  assume  an  initial--final  mass  relationship,  that is, a
relation  between  the  white  dwarf  mass and the  initial  mass of its
progenitor.  In Fig. 1 we show, as an  example,  the  location  of the
white dwarfs of an old cluster of initial solar  metallicity and ages of
10 and  12  Gyr,  respectively,  in  the  $M_{\rm  V}/(B-V)$  plane,  as
predicted  by  the  theoretical   models   (Salaris  et  al.~2000).  The
horizontal  lines mark the region  where  white  dwarfs  have  effective
temperatures  between  10\,000 and  20\,000 K, which is the  temperature
range of the template  field white dwarfs used by Renzini et al.  (1996)
to derive the distance to NGC~6752.

The initial--final mass relationship has been derived considering the CO
core masses after the first thermal pulse from the  evolutionary  models
of  Salaris  et  al.~(1997).  This   initial--final   mass  relationship
provides   white  dwarf  masses  nearly   constant  and  equal  to  0.54
$M_{\odot}$ for progenitor masses up to $\sim\,2.5\,M_{\odot}$, and then
increasing  up to 1.0  $M_{\odot}$  when  the  progenitor  mass  reaches
$7.0\,M_{\odot}$.  It can be  easily  seen in  Fig.~1  that the  cluster
white dwarf sequences show a pronounced turn to the blue at their dimmer
end  which,  as  the  age  of  the  cluster  increases,  is  located  at
increasingly  larger  magnitudes.  Until this blue--turn the white dwarf
sequence is almost  coincident  with the  cooling  track of the  $\sim\,
0.54\,  M_{\odot}$,  while the blue--turn is due to the  contribution of
more massive white dwarfs.  It is easy to understand  this  behaviour by
recalling  that at  each  brightness  along  the  cooling  sequence  the
constraint  $t_{\rm  GC}=t_{\rm  cool}+t_{\rm  prog}$  has to be  valid.
Since  $t_{\rm  cool}$ is very short at the  bright  end of the  cooling
sequence,  and  practically  negligible  with  respect to $t_{\rm  GC}$,
$t_{\rm prog}$ and, hence, the progenitor mass has to be, to a very good
approximation,  constant for a large magnitude  range, and very close to
the turn--off mass.  On the contrary, towards the dim end of the cooling
sequence $t_{\rm cool}$ becomes a sizeable fraction of $t_{\rm GC}$ and,
thus, the  contribution  of the white  dwarfs  coming  from  higher mass
progenitors  (and,  consequently,  with smaller $t_{\rm prog}$) is quite
apparent.  A  similar  feature  is  expected  for the halo  white  dwarf
population (Isern et al.  1998).

This feature can also be seen in Fig.~2  (upper  panel)  where we show
the run of the initial masses as a function of the $M_{\rm V}$ magnitude
along the cluster white dwarf sequences shown in Fig.~1.  In the $T_{\rm
eff}$  range we are  dealing  with  (marked,  as in Fig.~1, by  vertical
dashed  lines), the initial mass is almost  constant --- as well as, due
to our selected initial--final mass relationship, the actual white dwarf
mass  shown  in  the  lower  panel.  It  is  remarkable  that  even  two
magnitudes below the lower limit of this effective temperature range the
initial mass of white dwarf  progenitors is still constant.  This result
(the constancy of the progenitor  mass of white dwarfs  contributing  to
the effective  temperature  range, not its actual  value) is  completely
general,  that  is,  independent  of  the  adopted  initial--final  mass
relationship, since for all possible white dwarf masses the evolutionary
times are very fast at these high temperatures; moreover, this result is
also  independent  of  the  initial   metallicity  of  the  white  dwarf
progenitors  and  therefore  valid  also  in the  case  of  white  dwarf
progenitors  with lower  metallicities,  typical  of  galactic  globular
clusters.

Since the progenitor  mass is constant, one can then  reasonably  expect
that, even in the case of a  completely  different  initial--final  mass
relationship, the mean white dwarf mass and the spread around this value
is constant,  and equal to the values  attained at the  beginning of the
cooling sequence.  This behaviour is very different from what happens in
field disk white  dwarfs.  In this case,  because  of the  ongoing  star
formation  processes, very different white dwarf masses can populate the
same $T_{\rm eff}$ range, depending on the age of their progenitors.

There are in principle four quantities  which can introduce a systematic
error in the white dwarf distance  determination  to globular  clusters,
provided that there are significant  differences  between the properties
of the local template white dwarf sequence and the globular cluster one,
namely:

\begin{enumerate}
\item[i)]   The mass of the white dwarf cooling sequence.
\item[ii)]  The chemical composition of the envelope (DA, non--DA).
\item[iii)] The thickness (in mass) of the envelope.
\item[iv)]  The  chemical  stratification  of  the  core  (different  CO
	    profiles or, even, He--core white dwarfs).
\end{enumerate}

In the  following  we are going to discuss the  influence  of these four
parameters on the CMD location of white dwarf  cooling  sequences in the
$M_{\rm V}/(B-V)$ (hereinafter $BV$) and $M_{\rm V}/(V-I)$  (hereinafter
$VI$)  planes,  which are actually the most popular  CMDs  employed  for
studying  cluster  white  dwarfs --- see, e.g.,  Renzini et al.  (1996),
Richer et  al.~(1997),  and Von Hippel et  al.~(2000).  We consider as a
reference the $T_{\rm eff}$ range between 10\,000 and 20\,000 K, already
selected by Renzini et al.  (1996) for their distance determination.  In
this  temperature   range  the  white  dwarf  sequence  is  sufficiently
populated and bright enough to be detected with a reasonable photometric
error in close  globular  clusters,  using the HST or giant ground based
telescopes.

\subsection{The mass of the white dwarf sequence}

Stars in a globular  cluster  lose their mass along the Red Giant Branch
(RGB) due to stellar  winds.  The amount of mass lost during  this phase
is not the same for each red  giant  star,  and  this,  in  turn, is the
origin of the extended  horizontal  branches  (HB)  observed in Galactic
globular  clusters; this also means that the mass evolving along the RGB
(which at the  beginning  of the red giant phase is almost  equal to the
mass of the star at the  cluster  turn--off)  can give birth to HB stars
with  different  values of the total mass, but with the same  initial He
core mass.  These  different  HB stars  will all end up as white  dwarfs
with a mass range which is constrained  mainly by the size of the helium
core at the  beginning  of the  He-burning  phase  and by the mass  loss
processes along the subsequent Asymptotic Giant Branch (AGB) phase.

Renzini et al.  (1996) summarize the results about the few semiempirical
determinations  of white dwarf  masses in  galactic  globular  clusters,
providing a very narrow range of values,  namely  $0.53 \, \pm \, 0.02\,
M_{\odot}$.  This  range  of  values   nicely   overlaps   with  current
determinations  of the CO core mass at the first  thermal  pulse derived
from stellar  evolutionary models --- see, e.g.,  Wagenhuber~(1996)  and
Salaris  et  al.~(1997)  --- for  initial  masses  of  about  $0.8-1.0\,
M_{\odot}$.  Moreover, it is also in agreement  with  determinations  of
the mass distribution of nearby field white dwarfs, which peaks at about
$0.55  \,  M_{\odot}$  (Bragaglia  et  al.~1995;   Reid~1996).  However,
regarding  this last  result,  the reader  should  keep also in mind the
recent results of Bergeron, Leggett \& Ruiz (2000), who find a mean mass
of field white  dwarfs  about  $0.10 \,  M_{\odot}$  higher at lower $T_{\rm eff}$
values.

Very   recently   however,   Alves,   Bond  \&  Livio   (2000)   derived
semiempirically  the mass of the central  star in a planetary  nebula of
the  globular  cluster  M~15,  obtaining  a value  $0.60\,\pm\,  0.02 \,
M_{\odot}$.  It is not clear yet if this  result is an  indication  of a
globular  cluster   initial--final   mass  relationship  with  a  larger
dispersion  than that derived by Renzini et al.~(1996), or it is due (as
discussed  by the  authors)  to an  increment  of the  mass  through  an
interaction in a close binary system.  We just recall here that there is
also   another   possible   indication   of  a  large   dispersion   for
the initial--final  mass relationship in the results obtained by Reid (1996)
when considering the white dwarf  population of the Hyades open cluster.

\begin{figure}
\psfig{figure=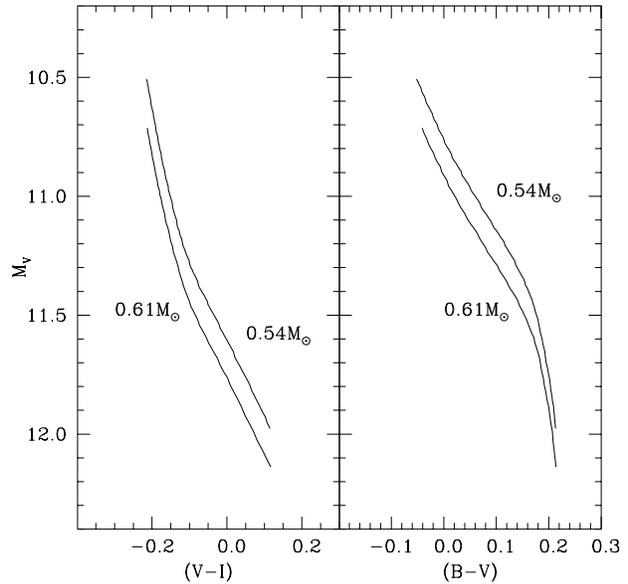,width=8.3cm,clip=}
\caption[]{White  dwarf  cooling  tracks in the $BV$ and $VI$ planes for
masses equal to 0.54 $M_{\odot}$ and 0.61 $M_{\odot}$, and $T_{\rm eff}$
between 20\,000 and 10\,000 K.}
\end{figure}

Another  important  piece of evidence  must also be taken into  account.
The determination of the mass range $0.53\, \pm\, 0.02 \, M_{\odot}$ ---
and the results of Alves et  al.~(2000),  as well --- is mainly based on
the maximum  brightness of AGB stars and on the  luminosity  of post-AGB
stars, which can be directly  related to the mass of the  degenerate  CO
core prior to the  beginning of the cooling  phase.  However,  there are
several globular clusters with an horizontal branch  morphology  showing
the  presence  of an  extended  blue  tail, as in the case of  NGC~6752.
Regardless  of the  physical  mechanism(s)  producing  the  stars  which
populate the hot side of the HB in globular  clusters,  it is well known
that they are stars  which  have lost a large  amount of their  envelope
during the RGB phase.  At the end of the He--burning  phase these stars,
depending on the mass of their residual H--rich  envelope, can behave as
post--Early  AGB  structures  or as  AGB-manqu{\'e}  ones  ---  see, for
instance,  Greggio  \&  Renzini  (1990)  --- and do not  experience  the
thermally  pulsing  phase on the AGB as most  massive  HB stars  do.  On
theoretical  grounds,  the  minimum  (initial)  HB mass  which  does not
experience the AGB thermal pulses is a quite robust  prediction  (Dorman
et al.  1993,  Bono et al.  1997),  being  equal to  $\approx  \, 0.52\,
M_\odot$.  Since  the mass of the CO core at the end of the HB  phase is
of the order of $0.45\, M_\odot$,  globular  clusters with extended blue
tails  can  produce   also  white   dwarfs  with  masses  in  the  range
$0.45-0.52\, M_\odot$, on average $\approx\,0.05\, M_\odot$ less massive
than the white dwarf progeny of stars  climbing up the AGB.  Thus, it is
worth  considering the  possibility of different masses for the globular
cluster cooling sequences.

Turning  now our  attention  to the CMD  location  of  white  dwarfs  of
different  masses, it is well  known  from  relatively  simple  physical
considerations  that the larger the white dwarf mass is, the smaller its
radius  is.  In  Fig.~3  we  display  in the $BV$ and $VI$  planes,  for
effective temperatures between 10\,000 and 20\,000 K, two representative
white  dwarf  cooling  tracks  with  masses  equal to 0.54 and  $0.61 \,
M_{\odot}$.  The higher mass is shifted to higher  $M_{\rm  V}$  values.
The derivative $\Delta M_{\rm V}/\Delta (M/M_{\odot})$ equals to 2.3 for
masses between 0.45 and 0.60 $M_{\odot}$.

\subsection{The chemical composition of the envelope}

As already  mentioned  in the  introduction,  field  white  dwarfs  with
$T_{\rm eff}$ between $\sim$ 20\,000 and 10\,000 K can be present either
in the DA or non--DA  types, with a number  ratio of DA to non--DA
equal to 4.
Our selected  standard  envelope  thicknesses (see next  subsection) are
$\log q({\rm H})=-4.0$ and $\log\-q({\rm He})=-2.0$ for DA white dwarfs,
and $\log q({\rm  He})=-3.5$ for the non--DA spectral type, where $q$(H)
and $q$(He)  indicate  the ratio of the mass  contained  in the H and He
envelope layers to the total white dwarf mass, respectively.

\begin{figure}
\psfig{figure=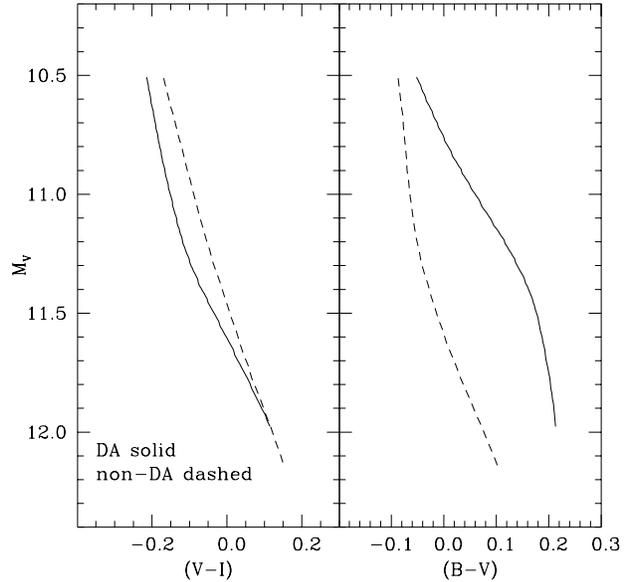,width=8.3cm, clip=}
\caption[]{Same  as Figure 3 but for a white  dwarf of 0.54  $M_{\odot}$
either DA or non--DA (see text for details).}
\end{figure}

The envelope  chemical  composition  strongly affects not only the white
dwarf cooling times, but also their location in the CMD, and the precise
shape of the cooling track, as clearly shown in Fig.~4 where we show the
cooling tracks for a $0.54 \, M_{\odot}$ DA white dwarf (solid line) and
a non--DA  white dwarf of the same mass.  It is evident that in the $BV$
plane DA white dwarfs are  brighter at a fixed  colour.  This is clearly
confirmed by observations of local DA white dwarfs, see e.g.  panel 2 of
Fig.~1 in Renzini et al.~(1996),  while the reverse  happens in the $VI$
plane.  Moreover, the separation  between DA and non--DA white dwarfs is
larger  in  the  $BV$  plane,  making   possibly   more  evident   their
identification in a given globular  cluster.  $M_{\rm V}$ differences of
up to $\sim$ 1 mag at a fixed $BV$ value are possible in the $BV$ plane,
while they are a factor  $\sim$ 5 smaller  in the $VI$  plane.  Also the
shape of the DA and non--DA cooling  sequences is different,  especially
in the $BV$ plane.

So far, we have considered DA and non--DA white dwarfs having their very
outer  layers made of either  pure  hydrogen  or pure  helium.  However,
observations  suggest that in this temperature range DA white dwarfs can
have a small  amount  of  helium  in  their  spectra,  of the  order  of
$10^{-5}$  by number, and that 20\% of non--DA  white  dwarfs  show some
hydrogen in their  envelopes, with  abundances of the order of $10^{-4}$
by number at most.  We have tested the effect of these small  admixtures
of helium in the outer  envelope of DA white  dwarfs, and of hydrogen in
the envelope of non--DA white  dwarfs, by computing  appropriate  models
for a $0.61  \,  M_{\odot}$  white  dwarf,  and  using  the  results  of
Bergeron,  Saumon  \&  Wesemael~(1995)  about  the  influence  of  H--He
mixtures on the derived white dwarf  colours.  We have found  negligible
variations of the CMD location  with respect to the case of our standard
He--free DA white dwarf envelopes and H--free non--DA atmospheres.

\subsection{The thickness of the envelope}

The thicknesses of the hydrogen and helium envelopes of white dwarfs has
been the  subject of many  investigations  during the last  decade.  The
thickness of these layers is a key ingredient to determine the evolution
of the white dwarf since, due to their opacity, they basically  regulate
the  energy  loss rate of the  isothermal,  highly  conductive  electron
degenerate  core.  Moreover,  the envelope  thickness  also  affects the
radius  of  the  white  dwarf  at  a  given  effective  temperature.  As
discussed by D'Antona \& Mazzitelli  (1990), due to the unknown  details
of the mass loss process  during the AGB phase and the planetary  nebula
ejection,  theoretical   evolutionary  models  cannot  yet  provide  too
stringent  predictions  about the  thickness of the hydrogen  and helium
layers surrounding the degenerate CO core.

Observational constraints based on spectroscopic (Bar\-stow et al.~1993)
as well as on asteroseismological (Cle\-mens~1995) analyses suggest that
local field DA white  dwarfs  have  typically  hydrogen  layers of about
$\log   q({\rm   H})=-4.0$.  On  the  other   hand,  the  study  of  the
mass--radius  relationship for a sample of field white dwarfs with known
Hipparcos  parallaxes  provides  indications  that the  thickness of the
hydrogen  layers spans a range of values  within  $\log  q({\rm  H})\sim
-4.0$ and $\log q({\rm  H})\sim -7.0$  (Provencal et  al.~1998).  As for
the  thickness  of the helium  layer  below the  hydrogen  envelope  the
assumed   reference  value  comes   basically  from  stellar   evolution
constraints,  and it is $\log q({\rm  He})\sim  -2.0$  (Hansen  1999 and
references therein).  We have verified, by computing white dwarf cooling
tracks with $\log q({\rm He})$  decreased  by 1 dex  (keeping  the total
white  dwarf mass  constant)  that the  location  in the CMD of DA white
dwarfs is basically  unchanged.  As for non--DA white dwarfs,  estimates
of the envelope  thickness range between $\sim 10^{-4}\,  M_{\odot}$ and
$\sim 10^{-2} \, M_{\odot}$  (Pelletier et al.~1986;  MacDonald, Hernanz
\& Jos\'e~1998).

\begin{figure}
\psfig{figure=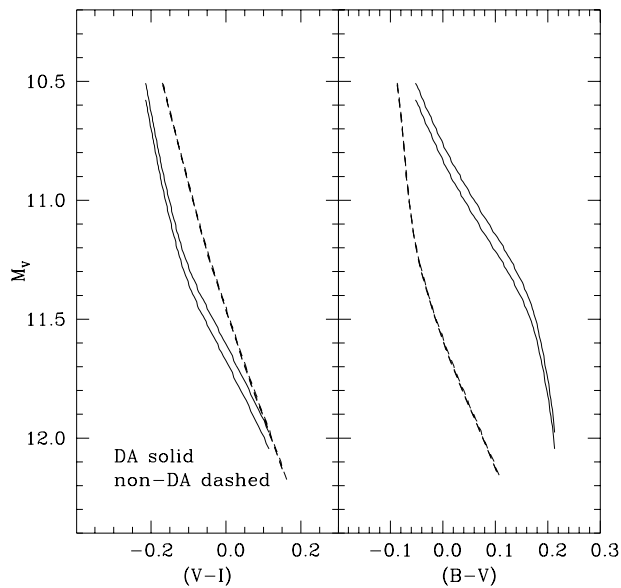,width=8.3cm,clip=}
\caption[]{Same as Figure 4 but for varying  thicknesses of the H and He
envelopes (see text for details).}
\end{figure}

All these  results are however for field  white  dwarfs; no  indications
exist yet about $\log  q({\rm H})$ and $\log  q({\rm  He})$ in  globular
clusters, apart from the fact that  theoretical  evolutionary  models of
white dwarfs coming from blue HB progenitors --- in the hypothesis  that
mass loss during the  He-burning  phases is  negligible  ---  predict an
upper limit to the  thickness  of the very outer H layers  ranging  from
$\log q({\rm H})\sim -3.5$ to $-4.0$ (Castellani et al.~1994a).

In Fig.~5 we show the effect of varying the  thickness of the external H
and He layers for,  respectively,  a DA and and a non--DA white dwarf of
$0.54\,  M_{\odot}$.  In the case of non--DA  white dwarfs our reference
value is $\log q({\rm He})=-3.5$, which we have changed by $\pm$1.0 dex,
obtaining  virtually no variation in the  location of the cooling  track
(the  three  tracks  perfectly  overlap  in  Fig.~5).  In the  case of H
envelopes  the  situation is  different,  since the  thickness  of the H
layers  affects  appreciably  the  location  of  the  track.  In  Fig.~5
representative  models  with $\log  q({\rm  H})=-4.0$  and $\log  q({\rm
H})=-6.0$ are  displayed; a reduction of the thickness of the H envelope
shifts  the track --- at a  constant  value of the  colour  ---  towards
higher $M_{\rm V}$ values, with a derivative  $\Delta  M_{\rm  V}/\Delta
\log q({\rm H}) \sim \, -0.035$ for $\log  q({\rm H})$  ranging  between
$-4.0$ and $-7.0$.

\subsection{The chemical stratification of the CO core}

The chemical  stratification  of the CO core may  potentially  affect as
well the CMD location of the white dwarf track since the the mass-radius
relationship  also depends on the electron mean molecular  weight of the
electron-degenerate  core.  To this regard, it is  important  to realize
that the value of the CO ratio  along the white dwarf core is subject to
some  uncertainties  due to our  poor  knowledge  of  the  value  of the
$^{12}{\rm  C}(\alpha,\gamma)^{16}{\rm  O}$  reaction  rate  --- see the
detailed  discussions in Salaris et al.~(1997) or Cassisi et al.~(2001),
and references  therein.  Moreover, for a given value of the  $^{12}{\rm
C}(\alpha,\gamma)^{16}{\rm O}$ reaction rate, the CO profile is possibly
a function of the  initial  metallicity  of the white  dwarf  progenitor
(Umeda et al.~1999).

\begin{figure}
\psfig{figure=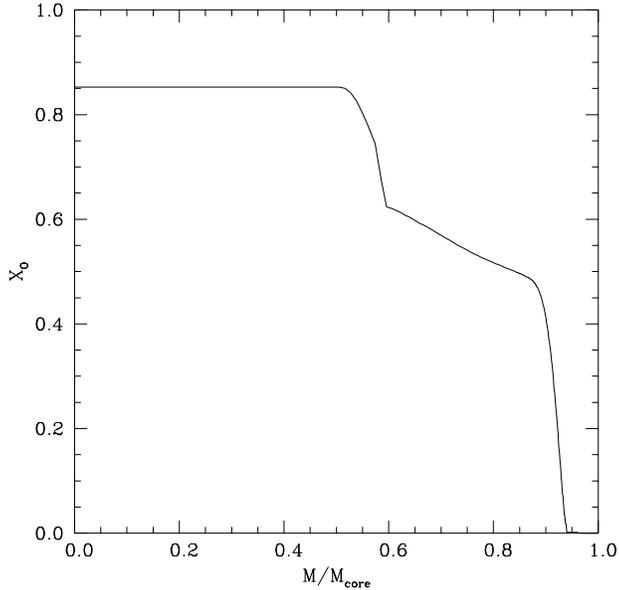,width=8.3cm,clip=}
\caption[]{Profile  of the oxygen mass fraction  ($X_{\rm O}$) along the
CO core of our reference  white dwarf model of $0.54 \, M_{\odot}$.  The
carbon abundance, $X_{\rm C}$, is $X_{\rm C}=1-X_{\rm O}$.}
\end{figure}

As a  numerical  test,  and in order to  mimic  these  effects,  we have
computed two cooling tracks for a $0.54\, M_{\odot}$ white dwarf, where,
for the first case, we have substituted our reference CO  stratification
shown in  Fig.~6  with a flat  profile  with  50\% of C and 50\% of O by
mass,  whereas in the second case we have  inverted the role of C and O,
preserving  the shape of the initial  chemical  profile.  The  resulting
cooling tracks are however  negligibly  affected by these changes in the
effective  temperature  range we are  dealing  with.  This  means  that,
regardless  of the still  large  uncertainty  affecting  the  $^{12}{\rm
C}(\alpha,\gamma)^{16}{\rm  O}$  reaction  rate, and  regardless  of the
possible   effects  on  the  CO  profile  of  the   initial   progenitor
metallicity,  the CMD location of the bright  portion of the white dwarf
cooling sequence is not appreciably affected.

\subsection{He--core white dwarfs}

He--core  white dwarfs are the  byproduct  of strong mass loss along the
red  giant  phase  (due for  example  to the  interaction  with a binary
companion),  which  strips  out the H  envelope  on top of the H burning
shell  before  the  degenerate  He core  reaches  the  critical  mass to
experience  the He flash and the  subsequent  quiescent He burning phase
(Kippenhahn, Kohl \& Weigert 1967;  Castellani,  Luridiana \& Romaniello
1994b; Hansen \& Phinney 1998).  Their mass can range from a value close
to the  degenerate  He  core--mass  at the red giant tip (about $0.50 \,
M_{\odot}$ for the most metal poor globular clusters as M~92, decreasing
down to about $0.48\, M_{\odot}$ for the metal rich ones as 47~Tuc) down
to masses of the  order of $0.20 \,  M_{\odot}$.  There is  possibly  at
least one direct  observational  indication  that He--core  white dwarfs
contribute  to the cooling  sequence of globular  clusters  (Moehler  et
al.~2000).

In Fig.~7 we show the CMD  location  of a $0.45 \,  M_{\odot}$  He--core
white dwarf with a H surface  layer of $\log  q({\rm  H})=-4.0$.  In the
same plot our  reference  DA and  non--DA  model  sequences  of $0.54 \,
M_{\odot}$  are also shown.  The  He--core  white dwarf mass is close to
the upper possible  value, lower masses being shifted to the red side of
the CMD.  As  expected  purely on the base of the mass  difference  with
respect to the DA white dwarf with a CO core  plotted in the figure, the
He--core  white  dwarf is  shifted  to  higher  brightnesses  at a fixed
colour.  This  feature is  reminiscent  of the two bright  white  dwarfs
discarded  by Renzini  et  al.~(1996)  in their  fitting  procedure  ---
compare the $BV$ panel in Fig.~7  with  Fig.~1 of Renzini et  al.~(1996)
--- because they were  clearly  located to the right side of the main DA
cluster  sequence.  In the $VI$ plane the He--core  white dwarf sequence
overlaps  with the non--DA  one at the bright end of the  $T_{\rm  eff}$
interval.  In  the  $BV$  plane,  due  to  the  steeper   slope  of  the
H--envelope  cooling  sequences, the He--core  white dwarf is on average
closer in colour to the CO one than in the $VI$ plane.

\begin{figure}
\psfig{figure=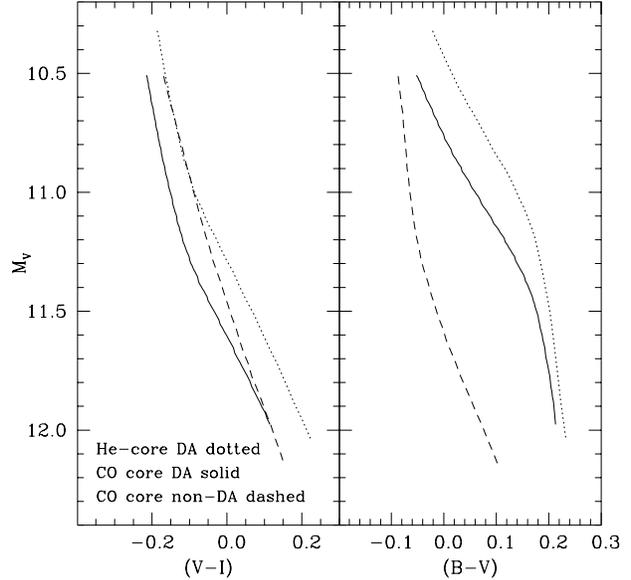,width=8.3cm,clip=}
\caption[]{Comparison  of the  CMD  of a  He--core  white  dwarf, with H
envelope and $M=0.45 \,  M_{\odot}$,  with our  reference DA and non--DA
sequences.}
\end{figure}

\section{Systematic errors on the WD fitting distances}

In this  section  we  quantify  the  possible  systematic  errors on the
distances derived from the WD--fitting technique taking into account the
results  previously  discussed.  We study the cases in which, due to our
lack of precise empirical or theoretical  determinations, the parameters
which  determine the position of the white dwarf sequence on the CMD are
possibly  different in the local  template and in the  globular  cluster
sequence,  but within the range  allowed  by  independent  observations.
Because the ratio  between the number of DA and non--DA  white dwarfs in
the  field is 4:1 (and  presumably  this  ratio is the same in  globular
clusters) one is forced to use the cooling  sequences of DA white dwarfs
in order to build the template sequence.  This, in turn, implies that an
important  role is played by the thickness of the H surface  layers.  To
this date, we do not have any  empirical  indication  about the value of
$q$(H) for white dwarfs in globular  clusters, and about its  dependence
on the HB morphology.  In principle, there could be sizeable  systematic
differences due, for example, to the fact that in globular clusters with
blue HB (like M~92 and NGC~6752)  the main  component of the white dwarf
population could be produced by stars not experiencing the AGB phase, in
contrast with white dwarfs  populating  red HB globular  clusters  (like
47~Tuc).  We are  forced,  therefore,  to  treat  the  thickness  of the
envelope layers as a free parameter, in the hypothesis that the possible
range of values is within the range spanned by field white dwarfs.

\subsection{Uncertainties on the template sequence}

Since the mass range for white dwarfs in globular clusters appears to be
mainly in the range $0.53\,\pm\,0.02\,  M_{\odot}$, it is clear that the
template white dwarf  sequence  should be made by stars within this mass
range.  However, there are almost no field white dwarfs suitable for the
WD--fitting  method,  for  which  the  value  of  the  mass  is  derived
empirically  (like,  for  instance,  the case of white  dwarfs  in close
binary  systems, as  Sirius~B).  In order to build the template DA white
dwarf  sequence,  one needs  therefore  to  employ  mass  determinations
obtained  from  semiempirical  methods, as  discussed  in  Bragaglia  et
al.~(1995).  It is possible to determine  spectroscopical values for the
gravity ($g$) and $T_{\rm  eff}$ of a sample of local white  dwarfs, and
then  derive the  masses  employing  theoretical  $\log  g-T_{\rm  eff}$
relationships.  However, the assumed  thicknesses of the envelope layers
of DA white dwarfs  affect the $\log  g-T_{\rm  eff}$  relationship.  In
fact, it can be shown that a uniform  distribution  of $\log q({\rm H})$
between  $-4.0$ and $-7.0$  produces a mass  range  dispersion  of about
$0.04 \,  M_{\odot}$  for a given  couple of $\log g$ and $T_{\rm  eff}$
values.

One could  speculate  if, after  deriving a certain  mass  range for the
template white dwarfs assuming a given spread in the $\log q$(H) values,
it is possible to judge if the  results  and the  assumptions  about the
envelope  thickness are consistent with the distribution of the stars in
the CMD.  As we are going to show  now,  this is not an easy  task  when
considering  realistic  errors  on  the  parallax  of  the  objects.  We
determined, using our cooling tracks, three template  sequences of about
100 DA white dwarfs by means of a simplified version of our Monte--Carlo
simulator  (Garc\'\i a--Berro et al.  1999), assuming a negligible error
on their  colours,  and  allowing  for an error  of  $\pm\,0.10$  mag in
$M_{\rm V}$, which is approximately  the average error of the brightness
of the template  white dwarfs of Renzini et al.~(1996)  due to the error
on their parallaxes  (which is of the order of 5\%).  The first sequence
(sequence A)  comprises  masses in the range  between  0.51 and $0.55 \,
M_{\odot}$, and $\log q$(H) uniformly distributed between $-4$ and $-7$;
sequence B has the same mass range but the value of the thickness of the
hydrogen outer layer was kept constant at $\log q({\rm H})=-4$; finally,
sequence C is  characterized  by $M=0.55 \, M_{\odot}$  and $\log q({\rm
H})=-4$.  In Fig.~8  we show the  three  sequences  in the $BV$ CMD (for
$T_{\rm eff}$ between 10\,000 and 20\,000 K), after shifting  vertically
sequence  B by +0.04  mag, and  sequence  C by  +0.015  mag, in order to
reproduce the average brightness of sequence A.  The average  brightness
of the three  sequences  is  different,  but the  dispersion  around the
average looks very much the same, and  dominated by the parallax  error,
in spite of the fact that the mass and/or envelope  thickness ranges are
different.

The main result of this exercise is that  uncertainties  in the value of
$\log q$(H) produce  unavoidable  uncertainties  on the precise value of
the  masses of the  template  white  dwarfs  and,  thus, on the  precise
location of the template  sequence,  which may  contribute  to the error
budget of the WD--fitting technique by amounts of the order of less than
0.05  mag.  For the  sake  of  conciseness,  in the  following  we  will
consider  as our  reference  template a  sequence  (determined  from the
Monte--Carlo  simulations) made of about 100 DA white dwarfs with masses
in the range between 0.51 and $0.55 \, M_{\odot}$, $\log q$(H) uniformly
distributed  between  $-$4 and $-$7 and  effective  temperatures  in the
range between 10\,000 and 20\,000 K.

\begin{figure}
\psfig{figure=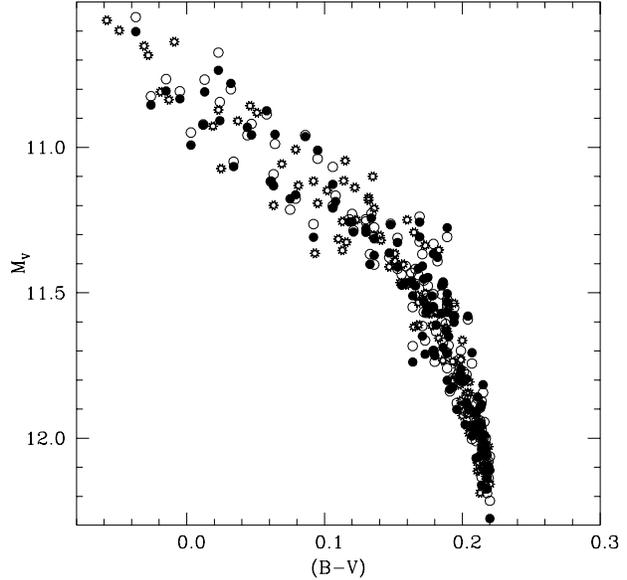,width=8.3cm,clip=}
\caption[]{  Comparison  among   a white dwarf  template  sequence  with
masses between 0.51 and $0.55 \, M_{\odot}$ and $\log q$(H) between $-4$
and $-7$  (sequence  A, filled  circles), a sequence  with the same mass
range but  $\log  q({\rm  H})=-4$  (sequence  B, empty  circles),  and a
sequence with $M=0.55 \, M_{\odot}$ and $\log q({\rm  H})=-4$  (sequence
C, starred  circles).  All sequences  have been  calculated  including a
random error of 0.10 mag in their  $M_{\rm V}$  magnitudes.  Sequences B
and C have been also shifted vertically by, respectively, 0.04 and 0.015
mag in order to overlap with sequence A.}
\end{figure}

\subsection{Uncertainties   on  the  white   dwarf  mass  and   envelope
thickness.  Clusters with a red HB}

We define here as red HB globular  clusters those  clusters in which the
mass of the HB stars is always larger than $\sim \, 0.52 \,  M_{\odot}$,
corresponding  to a colour of HB stars  larger than  $(B-V)\sim  -$0.21.
This means that all the white dwarfs  populating  red HB clusters  are a
product of AGB evolution (see the  discussion in our previous  section).
In this case, empirical estimates of the white dwarf masses show a range
between 0.51 and $0.55 \, M_{\odot}$.  Thus, if the unknown thickness of
the H layers of the DA cluster  white  dwarfs is similar  to that of the
local white  dwarfs, in  principle  there  should be no  systematic
error in the distances derived using our reference template sequence.

To obtain a  reasonable  estimate  of the  maximum  possible  systematic
error, our  reference  sequence has been fitted to a cluster DA sequence
with, respectively,  $M=0.51\,  M_{\odot}$ and $\log q({\rm H})=-4$, and
$M=0.55 \, M_{\odot}$ and $\log q({\rm H})=-7$.  The systematic error in
the distance  modulus is of $-0.10$ mag in the first case and of $+0.10$
mag in the second  case,  with a  $\sim\,60$\%  contribution  due to the
effect of the envelope  thickness.  It is obvious  that in the case that
the cluster white dwarfs have an  intrinsic  spread in mass
and/or  $\log  q({\rm H})$ within the  mentioned  range, the  systematic
error on their distance modulus is smaller than $\pm\,0.10$ mag.

If one accepts the results of Alves et al.~(2000) about M~15 as due to a
genuine spread in the initial--final  mass  relationship, the mass range
of globular  cluster  white dwarfs could  possibly  extend up to $0.60\,
M_{\odot}$.  By  repeating  the  previous  exercise  considering  as  an
extreme case  $M=0.60 \,  M_{\odot}$  and $\log  q({\rm  H})=-7$ for the
cluster  white  dwarfs, one gets  systematic  errors of +0.20 mag in the
cluster  distance  modulus.  Of  course  this  is a very  extreme  case.
Probably, it si more realistic to consider a mass  distribution  between
0.51 and $0.60 \, M_{\odot}$ and $\log q({\rm H})$  distributed  between
$-4$ and $-7$.  With this sample of 100 cluster  white  dwarfs and using
the same  reference  template  white  dwarf  sequence,  we  determine  a
systematic  error on the distance  modulus of +0.05 mag.  Should all the
globular  cluster  white  dwarfs have the same H layers  thickness,  the
systematic  errors would be then $+0.01$ mag if $\log q({\rm H})=-4$ and
$+0.11$ mag if $\log q({\rm H})=-7$.

\subsection{Uncertainties   on  the  white   dwarf  mass  and   envelope
thickness.  Clusters with a blue HB}

We define as blue HB globular  clusters those clusters in which the mass
of the HB stars extends below  $\sim\,  0.52\,  M_{\odot}$.  In general,
even  globular  clusters  with long blue tails always have a fraction of
their HB population  located at colours  larger than  $(B-V)\sim  -0.21$,
and, therefore, white dwarfs populating  clusters with a blue HB are the
product of both AGB evolution and  AGB-manqu\'e or post-Early AGB stars.
As discussed  before, one expects  that this kind of globular  clusters
is  populated  by white  dwarfs  with  masses  down to $\sim  \,0.45 \,
M_{\odot}$.  It appears that blue HB clusters  should  therefore  show a
larger mass  spread for the white  dwarf  population,  with white  dwarf
masses ranging between approximately 0.45 and $0.55 \, M_{\odot}$.

\begin{figure}
\psfig{figure=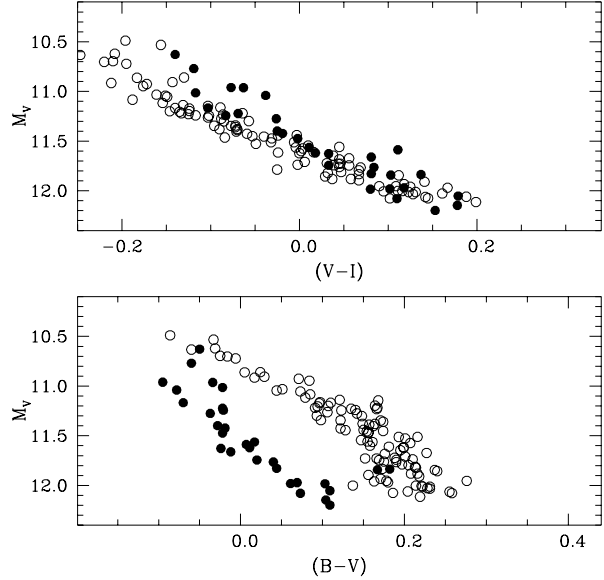,width=8.3cm,clip=}
\caption[]{$BV$  and  $VI$  colour--magnitude  diagrams  for  DA  (empty
circles)  and  non--DA  (filled  circles)  white  dwarfs  (see  text for
details) assuming observational errors of 0.02 mag in $B$, $V$ and $I$.}
\end{figure}

\begin{figure}
\psfig{figure=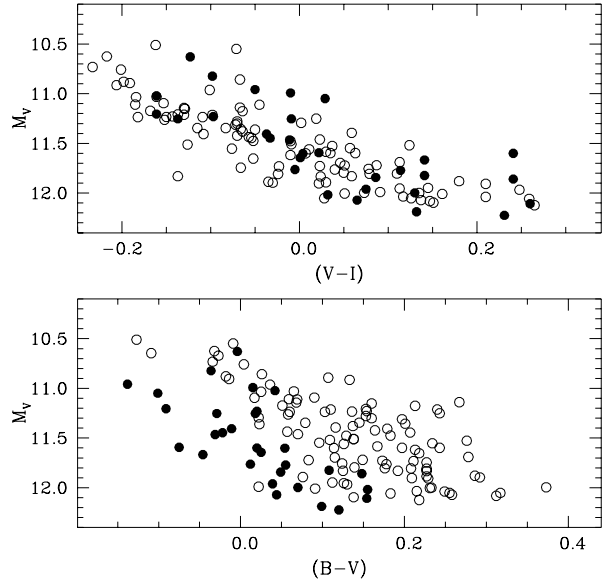,width=8.3cm,clip=}
\caption[]{Same as in Figure~9 but for observational  errors of 0.05 mag
in $B$, $V$ and $I$.}
\end{figure}

By fitting our template sequence to a cluster sequence with a mass range
between  0.45 and $0.55 \,  M_{\odot}$  and a range of $\log q$(H) as in
the template one, we obtained a systematic error on the distance modulus
of  $-$0.06  mag.  If we  accept a larger  mass  range of up to  $0.60\,
M_{\odot}$,  the  systematic  error goes down to $-0.04$  mag, since the
higher mass objects tend to compensate  for the presence of masses lower
than the  template  ones.  This is of  course  just an  estimate  of the
average  systematic  error in case of globular  clusters with a blue HB.
To obtain more precise evaluations one should compute synthetic CMDs for
each  given  cluster  in  order  to  determine,  from  the  observed  HB
morphology,  the mass  distribution  along the HB and  therefore, on the
base of the  discussion in Sect. 3.1, infer the possible  distribution
of white dwarf masses.

Considering  now  the  possibility  of  different  envelope  thicknesses
between template and cluster white dwarfs, if the white dwarf mass range
is between 0.45 and $0.55 \, M_{\odot}$ and $\log q({\rm  H})=-4$ in the
cluster,  the  systematic  error on the derived  distance  modulus is of
about $-0.12$ mag.  If we adopt the same mass distribution but we change
$\log  q$(H) to $-7$,  the  error is  negligible,  since  the  different
envelope thickness  compensates for the different mass range between the
template and cluster sequence.

\subsection{Contamination of the DA sample by non--DA white dwarfs}

Up to know, we have made the assumption that cluster DA white dwarfs can
be  distinguished  from  non--DA  ones.  In absence  of  spectroscopical
identification,  the only way to  discriminate  between  DA and  non--DA
white dwarfs is by checking the relative CMD location of the white dwarf
sample (see  Fig.~4).  Here we estimate  the maximum  photometric  error
allowing for a clear  distinction  between DA and non--DA objects in the
$BV$  and  $VI$  CMDs.  Moreover,  we  estimate  the  systematic  errors
introduced in the globular  cluster  distances  when using a DA template
sequence  fitted to a cluster  white  dwarf  sample  made of both DA and
non--DA  stars.  We assume that the cluster  DA/non--DA  number ratio in
the $T_{\rm  eff}$  interval we are dealing with is 4:1 as in the field.
Notice,  however,  that  below  $T_{\rm  eff}=10\,000$~K  the  number of
non--DAs  is similar or even  larger  than the  number of DAs.  The most
reasonable  explanation  for this is that the  outer  convective  region
mixes the H layer into the He envelope (Bergeron et al. 2000).

Figs.~9  and 10 show a  sequence  of about  80 DA and 20  non--DA  white
dwarfs  ($M=0.54 \,  M_{\odot}$,  $\log  q({\rm  H})=-4$ for the DAs and
$\log q({\rm He})=-3.5$ for the non--DAs) computed  including  1$\sigma$
photometric errors in $B$, $V$, and $I$ of, respectively,  0.02 and 0.05
mag.  Even in the  case  of the  smallest  errors,  the  non--DA  and DA
sequences  are not  well  separated  in the  $VI$  plane,  while  the DA
sequence is clearly  distinguishable from the non--DA one in the $BV$ plane.
With an error bar of the order of 0.05 mag, non--DA  white  dwarfs start
to be mixed up with DA ones also in the $BV$ plane.  In order to be more
precise,  when  fitting a template DA sequence to a cluster  white dwarf
sequence including both DA and non--DA objects, one  underestimates  the
globular  cluster  distance by only 0.03 mag if the fitting is performed
in the $VI$ plane, whereas in the case in which the fitting is performed
in the $BV$ plane the distance is  overestimated  by $\sim\,  0.20$ mag.
These figures are of course  reduced if the ratio  DA/non--DA is smaller
than the value we have adopted here.

\subsection{He-core white dwarfs}

Another  potential  source  of  systematic  errors  in  the  WD--fitting
distances is the presence of He--core  white  dwarfs (see  Fig.~7),  and
although it is hard to give a precise  quantitative  assessment  of this
error  source, we can  however  provide an  argument  to give some hints
about the  magnitude of this effect.  We will show that the effect would
be possibly  small or even  negligible.  White dwarfs with cores made of
He are the product of binary  evolution.  Actual estimates of the binary
frequency in globular  clusters  provide a value of the order of 10\% or
less (Hut et al.~1992).  If one accepts this  estimate, we have verified
that, by  computing  a cluster  cooling  sequence  containing  about 130
CO--core  white dwarfs with  DA/non-DA  ratio 4:1 and constant  envelope
thickness ($\log q({\rm H})=-4$ for the DAs and $\log q({\rm  He})=-3.5$
for the  non--DAs),  plus  about 14  (about  10\% of the  total  sample)
He--core DA objects,  even if all the He--core  white dwarfs have masses
of the order of $0.45 \, M_{\odot}$,  their presence causes a systematic
error of the order of only  $-$0.01  mag.  In the case  that  there is a
spectrum of masses for He--core white dwarfs,  ranging  between $\sim \,
0.20$ and $0.45 \,  M_{\odot}$,  then even  less  objects  will be close
enough  to  the  CO  cooling   sequences  for  the  relevant   effective
temperature range to be confused with them, while the others will lay at
larger colours,  clearly  separated from the more populated CO sequences
(see Fig.~11).

\begin{figure}
\psfig{figure=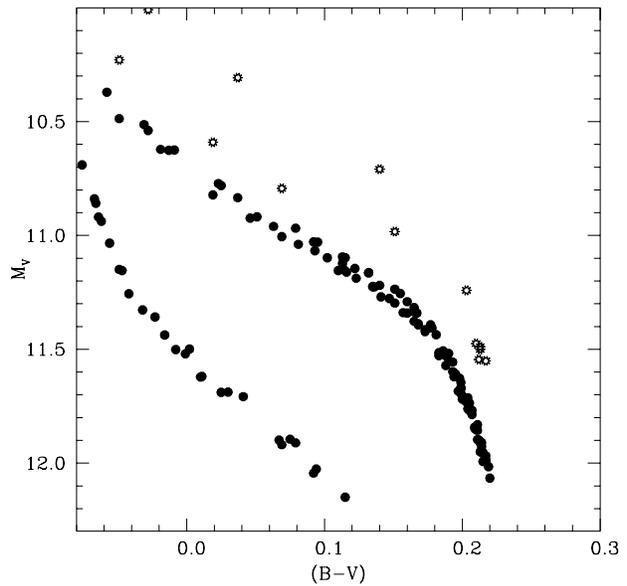,width=8.3cm,clip=}
\caption[]{CMD  of a globular cluster cooling  sequence made of CO--core
white dwarfs with masses  between 0.51 and 0.55  $M_{\odot}$,  DA/non-DA
ratio 4:1 (filled  circles),  plus a 10\%  fraction of He--core DA white
dwarfs  with  masses  uniformly   distributed   between  0.20  and  0.45
$M_{\odot}$  (starred  circles).  For all objects the envelope thickness
is  constant  and equal to $\log  q({\rm  H})=-4$  for the DAs and $\log
q({\rm He})=-3.5$ for the non--DAs.}
\end{figure}

\subsection{Cluster reddening}

The knowledge of the reddening of the globular cluster is fundamental in
order to derive the cluster  distance using the  WD--fitting  technique.
We tested the sensitivity of the derived distances to the uncertainty on
the cluster reddening, by considering two DA cooling tracks with $M=0.54
\,  M_{\odot}$  and $\log  q({\rm  H})=-4$,  shifted in colour  one with
respect to the other; we then fitted the two sequences one of top of the
other for different amounts of the relative shift, obtaining in the $BV$
and $VI$  planes a value for the  derivative  of the  apparent  distance
modulus with respect to the $(B-V)$ colour  excess,  $\Delta  (m-M)_{\rm
V}/\Delta E(B-V)\,\sim\,-5.5$.

\section{Summary and conclusions}

In this paper we have  thoroughly  discussed  many  possible  sources of
systematic errors on the globular clusters distances  obtained using the
WD--fitting  technique.  In order to do so we have first investigated in
detail the changes  produced in the $BV$ and $VI$ CMDs when  varying the
adopted mass of the template white dwarf cooling sequence, the influence
of the assumed  thicknesses  of their  envelopes on the  location of the
cooling  tracks, as well as the  consequences  of changing  the chemical
composition of their envelopes and of their cores.  We have then applied
these results in order to derive, by means of Monte--Carlo  simulations,
a  realistic   estimate  of  the  systematic   errors  involved  in  the
application of this technique to galactic  globular  clusters.  Our main
results can be summarized as follows:

\begin{enumerate}

\item The unknown  thickness  of the H layers in cluster DA white dwarfs
plays  a  non--negligible   role,  comparable  to  the  role  played  by
uncertainties  on the white dwarf  masses.  For  reasonable  assumptions
(derived from  observations of field white dwarfs and  constraints  from
globular  cluster  CMDs) about the  unknown  mass and $\log  q({\rm H})$
values in cluster DA white  dwarfs, a realistic  estimate of the maximum
systematic  error on the derived  distance moduli is within $\pm\, 0.10$
mag.  However, one should be aware that particular combinations of white
dwarf masses and envelope  thicknesses --- still allowed by current weak
or  non-existent  observational  constraints  --- could  produce  larger
errors.

\item A photometric  precision better than $\sim\,0.05$ mag is needed in
order to distinguish DA from non--DA white dwarfs in the $BV$ plane.  An
even better  precision is needed when using the $VI$ plane.  In the case
of  larger   observational   errors   and  no  clear   (spectroscopical)
distinction  between DA and  non--DA  cluster  white  dwarfs,  fitting a
template DA sequence to a cluster  sequence  made of a mixture of DA and
non--DA stars introduces a very small systematic error $\sim\,-0.03$ mag
in the $VI$ plane, but this error amounts to  $\sim\,+0.20$  mag, in the
$BV$  plane.  The $VI$  plane  looks  therefore  better  suited  for the
application  of the  WD--fitting  method, since it permits to get rid of
the  uncertainty  due to a  possible  contamination  of the DA sample by
non--DA white dwarfs.

\item  Contamination  by  He--core  white  dwarfs  should not  influence
appreciably the WD--fitting distances.

\item Due to the steep slopes of the white dwarf  cooling  curves in the
CMD, the  distance  derived  from the  WD--fitting  technique  has a non
negligible  dependence on the adopted cluster  reddening.  We obtained a
derivative  $\Delta(m-M)_{\rm V}/\Delta  E(B-V)\sim\,-5.5$, a dependence
which is similar  to the one of the  MS--fitting  technique  in the $BV$
CMD.

\end{enumerate}

\begin{acknowledgements}  We thank M.  Zoccali for useful  comments to a
preliminary  version of the manuscript.  Part of this work was supported
by the Spanish DGES project  numbers  PB98--1183--C03--02,  AYA2000-1785
and ESP98-1348, by the CIRIT and by Sun MicroSystems  under the Academic
Equipment  Grant  AEG--7824--990325--SP.  One  of  us  (S.C.)  has  been
supported  by  MURST-Cofin2000- under the scientific project
"Stellar Observables of Cosmological Relevance".

\end{acknowledgements}

{}

\end{document}